\documentclass[sigconf]{acmart}

\acmConference[ICSE 2022]{The 44th International Conference on Software Engineering}{May 21–29, 2022}{Pittsburgh, PA, USA}

\copyrightyear{2022} 
\acmYear{2022} 
\setcopyright{acmlicensed}\acmConference[ICSE '22]{44th International Conference on Software Engineering}{May 21--29, 2022}{Pittsburgh, PA, USA}
\acmBooktitle{44th International Conference on Software Engineering (ICSE '22), May 21--29, 2022, Pittsburgh, PA, USA}
\acmPrice{15.00}
\acmDOI{10.1145/3510003.3510047}
\acmISBN{978-1-4503-9221-1/22/05}

\newif\ifDEBUG
\DEBUGfalse

\newif\ifBLINDED
\BLINDEDfalse

\newcommand{\code}[1]{\texttt{{\small #1}}}

\newcommand{\ie}{\textit{i.e.,\ }}
\newcommand{\eg}{\textit{e.g.,\ }}
\newcommand{\etal}{\textit{et al.\ }}
\newcommand{\etals}{\textit{et al.}'s\ }

\usepackage{mdframed}
\mdfsetup{skipabove=0.5\topskip,skipbelow=0.5\topskip,align=center}

\usepackage{amsthm}
\newtheorem{thm}{Theorem}

\newcommand{\REDOS}{ReDoS\xspace}

\newcommand{\KREsAcros}{K-regexes\xspace}

\usepackage{enumitem}
\setlist[itemize]{leftmargin=*}
\setlist[enumerate]{leftmargin=*}

\usepackage{todonotes}
\usepackage{soul}

\ifDEBUG
    \newcommand{\DA}[1]{\todo[color=white,inline]{DA:#1}}
    \newcommand{\EB}[1]{\todo[color=green,inline]{EB:#1}}
    \newcommand{\XD}[1]{\todo[color=red,inline]{XD:#1}}
    \newcommand{\JD}[1]{\todo[color=yellow,inline]{JD:#1}}
    
    \newcommand{\TODO}[1]{\hl{#1}}
\else
    \newcommand{\DA}[1]{}
    \newcommand{\EB}[1]{}
    \newcommand{\XD}[1]{}
    \newcommand{\JD}[1]{}
    
    \newcommand{\TODO}[1]{}
\fi

\usepackage{booktabs} 
\usepackage{colortbl}
\usepackage{amsmath}
\usepackage{amsfonts}
\usepackage{algpseudocode}
\usepackage[normalem]{ulem} 
\usepackage{xspace}
\usepackage{multirow} 
\usepackage{balance} 
\usepackage{fancyvrb} 
\usepackage[export]{adjustbox} 
\usepackage{subcaption} 
\usepackage[font=normalsize,skip=5pt]{caption}
\usepackage{url}
\usepackage{xcolor}

\usepackage{tikz} 
\usetikzlibrary{automata} 
\usetikzlibrary{positioning} 
\usetikzlibrary{arrows} 
\usetikzlibrary{calc} 
\usetikzlibrary{patterns} 
\usetikzlibrary{snakes} 

\usepackage[group-separator={,}]{siunitx}
\usepackage{stfloats}
\VerbatimFootnotes 

\usepackage{cleveref}
\crefformat{section}{\S#2#1#3}
\crefname{figure}{Figure}{Figures}
\crefname{table}{Table}{Tables}
\crefname{listing}{Listing}{Listings}
\crefname{theorem}{Theorem}{Theorems}
\crefname{thm}{Theorem}{Theorems}
\crefname{lemma}{Lemma}{Lemmata}
\crefname{equation}{Eqt.}{Eqts.}
\crefformat{Grammar}{Grammar #1}

\usepackage{listings}

\newcommand{\myparagraph}[1]{\paragraph{#1}}
\renewcommand{\myparagraph}[1]{\vspace{0.25em} \noindent \textbf{#1}}

\newenvironment{RQList}{
   \setlength{\topsep}{0pt}
   \setlength{\partopsep}{0pt}
   \setlength{\parskip}{0pt}
   \begin{description}[style=unboxed]
   \setlength{\leftmargin}{1in}
   \setlength{\parsep}{0pt}
   \setlength{\parskip}{0pt}
   \setlength{\itemsep}{0pt}
   }
   {\end{description}}

\usepackage{newfloat}
\usepackage{syntax}
\usepackage{framed}

\DeclareFloatingEnvironment[
  fileext   = logr,
  listname  = {List of Grammars},
  name      = Grammar,
  placement = htp
]{Grammar}

\newcommand{\Wustholz}{{W{\"u}stholz} }

\usepackage[ruled,vlined]{algorithm2e}

\newcommand{\FindingsBox}[2]{
    \begin{mdframed}[backgroundcolor=black!10]
    \textbf{Finding #1:}
    #2
    \end{mdframed}
}

\newcommand{\ApisGuruDomainCount}{475\xspace}

\newcommand{\ApisGuruSubdomainCount}{1,059\xspace}

\newcommand{\APIRegexValidatingDocuments}{549\xspace}

\newcommand{\APIRegexUsingDomainPercent}{17.4\%\xspace}

\newcommand{\APIRegexUsingSubdomainPercent}{30.4\%\xspace}

\newcommand{\APITotalRegexUseNumber}{2681\xspace}
\newcommand{\APIUniqueRegexNumber}{1841\xspace}

\newcommand{\APIRegexUsingDomainCount}{83\xspace}
\newcommand{\APIRegexUsingSubdomainCount}{322\xspace}

\newcommand{\SLRegexDomainCount}{17\xspace}

\newcommand{\SLRegexDomainCountAPI}{11\xspace}

\newcommand{\SLRegexSubdomainCount}{45\xspace}

\newcommand{\SLRegexSubdomainCountAPI}{41\xspace}

\newcommand{\NProbedAPIDomains}{10\xspace}
\newcommand{\NProbedAPISubdomains}{32\xspace}
\newcommand{\UnsafeDomains}{6\xspace}
\newcommand{\UnsafeSubdomains}{15\xspace}

\newcommand{\ConclusivelySafeSubdomains}{16\xspace}

\newcommand{\SLDomainsInTopThousand}{2\xspace}
\newcommand{\AWS}{\emph{A}\xspace}
\newcommand{\Beezup}{\emph{B}\xspace}
\newcommand{\Azure}{\emph{C}\xspace}
\newcommand{\Tomtom}{\emph{D}\xspace}
\newcommand{\Monvoyage}{\emph{E}\xspace}
\newcommand{\Here}{\emph{F}\xspace}
\newcommand{\ImageCharts}{\emph{G}\xspace}

\newcommand{\SwaggerCodeGenNStars}{14K\xspace}
\newcommand{\SwaggerCodeGenNContributors}{1K\xspace}
\newcommand{\OpenAPIGeneratorNStars}{11.2K\xspace}
\newcommand{\OpenAPIGeneratorNContributors}{2K\xspace}

\newcommand{\WebFormsSampleSize}{1,000\xspace}

\newcommand{\WebFormsTopXPopulationSize}{1 million\xspace}

\newcommand{\ApifyMaxCrawlDepth}{500\xspace}
\newcommand{\ApifyPercentFullyCrawled}{66\%\xspace}

\newcommand{\WebFormRegexDomainPercentHasForm}{80\%\xspace} 
\newcommand{\WebFormRegexUsingDomainCount}{272\xspace} 
\newcommand{\WebFormRegexUsingDomainPercent}{39.1\%\xspace}
\newcommand{\WebFormRegexUsingPageCount}{30895\xspace} 
\newcommand{\WebFormRegexUsingPagePercent}{20.6\%\xspace}

\newcommand{\NumOfDomainUsePattern}{31\xspace} 
\newcommand{\NumOfDomainUseRegexFromJS}{265\xspace} 
\newcommand{\NumOfDomainUseRegex}{272\xspace} 

\newcommand{\NumTotalServices}{355\xspace}
\newcommand{\NumTotalUnsafeRegex}{17\xspace}

\newcommand{\NumOfDomainAtLeastOnePage}{696\xspace}

\newcommand{\MediumNumPagePerDomain}{104\xspace} 
\newcommand{\MediumNumFormPerDomain}{33\xspace} 

\newcommand{\WebFormTotalPatternNumber}{4966\xspace}
\newcommand{\WebFormUniquePatternNumber}{33\xspace}

\newcommand{\VulnRegex}{6\xspace}
\newcommand{\DomainWithVulnRegex}{6\xspace}

\newcommand{\DomainWithVulnRegexWithinDomainWithRegex}{2\%\xspace}

\newcommand{\PercentServicesRejectingGCPConnections}{3.4\%\xspace}

\newcommand{\ProbeNWarmupRequests}{3\xspace}
\newcommand{\ProbeNTreatmentAndControlRequests}{5\xspace}

\newcommand{\NInconclusiveResultsAcrossAllInterfaces}{12\xspace}

\newcommand{\ClassificationRSQthreshold}{0.25\xspace}

\begin{document}

\title{Exploiting Input Sanitization for Regex Denial of Service}

\author{Efe Barlas}
\authornote{Both authors contributed equally to this research.}
\affiliation{%
  \institution{Purdue University}
  \city{West Lafayette}
  \state{Indiana}
  \country{USA}
  \postcode{47906}
}
\email{ebarlas@purdue.edu}

\author{Xin Du}
\authornotemark[1]
\affiliation{%
  \institution{Purdue University}
  \city{West Lafayette}
  \state{Indiana}
  \country{USA}
  \postcode{47906}
}
\email{du201@purdue.edu}

\author{James C. Davis}
\affiliation{%
  \institution{Purdue University}
  \city{West Lafayette}
  \state{Indiana}
  \country{USA}
  \postcode{47906}
}
\email{davisjam@purdue.edu}

\begin{abstract}
  Web services use server-side input sanitization to guard against harmful input. 
  Some web services publish their sanitization logic to make their client interface more usable, \eg allowing clients to debug invalid requests locally.
  However, this usability practice poses a security risk.
  Specifically, services may share the regexes they use to sanitize input strings --- and regex-based denial of service (\REDOS) is an emerging threat.
  Although prominent service outages caused by \REDOS have spurred interest in this topic, we know little about the degree to which live web services are vulnerable to \REDOS.
  
  In this paper, we conduct the first black-box study measuring the extent of \REDOS vulnerabilities in live web services.
  We apply the \emph{Consistent Sanitization Assumption}: that client-side sanitization logic, including regexes, is consistent with the sanitization logic on the server-side.
  We identify a service's regex-based input sanitization in its HTML forms or its API, find vulnerable regexes among these regexes, craft \REDOS probes, and pinpoint vulnerabilities.
  We analyzed
    the HTML forms of \WebFormsSampleSize services 
  and
    the APIs of \ApisGuruDomainCount services.
  Of these,
    \NumTotalServices services publish regexes;
    \NumTotalUnsafeRegex services publish unsafe regexes;
    and
    6 services are vulnerable to \REDOS through their APIs (6 domains; 15 subdomains).
  Both Microsoft and Amazon Web Services patched their web services as a result of our disclosure.
  Since these vulnerabilities were from API specifications, not HTML forms,
  we proposed a \REDOS defense for a popular API validation library, and our patch has been merged.
  To summarize: in client-visible sanitization logic, some web services advertise \REDOS vulnerabilities in plain sight.
  Our results motivate short-term patches and long-term fundamental solutions.

{
  \vspace{5pt}
  
  \textit{``Make measurable what cannot be measured.'' \hfill --Galileo Galilei} 
}
  
\end{abstract}



\maketitle

 
\vspace{-0.09cm}
\section{Introduction} \label{section:Introduction}


Internet-based web services play a major role in modern society.
By their nature, web services are accessible through an interface, and so they must handle input from users both legitimate and adversarial.
Web services interpret string-based inputs into appropriate types such as email addresses, phone numbers, and credit card information.
A common first line of defense is therefore to filter for reasonable-looking input. 
If this input sanitization is flawed, the health of the web service can be compromised~\cite{SomeSoKAndCSURAboutInputSanitization}.

Unfortunately, a common input sanitization strategy exposes web services to a denial of service attack called Regular expression Denial of Service (\REDOS).
Many software systems rely on regular expressions (regexes) for input sanitization~\cite{Chapman2016RegexUsageInPythonApps,Davis2019RegexGeneralizability}.
Some of these regexes are \emph{problematically ambiguous}~\cite{Brabrand2010DisambiguatingRegexes,Weideman2016REDOSAmbiguity} and may require super-linear time (in the input length) to evaluate in an unsafe regex engine~\cite{Thompson1968LinearRegexAlgorithm}.
At present, most regex engines are unsafe in this regard~\cite{Cox2007RegexMatchingCanBe,Davis2019LinguaFranca}.
The cost of regex processing, combined with the use of regexes across the system stack, can affect the availability of web services, leading to regex-based denial of service (\REDOS)~\cite{Crosby2003REDOS,Crosby2003AlgorithmicComplexityAttacks}.

The \REDOS problem has been considered from several perspectives.
Theoretically, the properties of problematic regexes under different search models have been established,
  including both Kleene-regular semantics~\cite{Rathnayake2014rxxr2,Weideman2016REDOSAmbiguity,Wustholz2017Rexploiter}
  and 
  extended semantics~\cite{liuRevealerDetectingExploiting2021}.
In terms of the supply chain, Davis \etal showed that up to 10\% of the regexes in open-source modules are problematically ambiguous~\cite{Davis2018EcosystemREDOS,Davis2019LinguaFranca,Davis2019RegexGeneralizability}.
With respect to live services,
  \Wustholz \etal showed that problematic regexes in many Java applications are exploitable~\cite{Wustholz2017Rexploiter},
  and Staicu \& Pradel showed that 10\% of the Node.js-based web services they examined were vulnerable to \REDOS~\cite{Staicu2018REDOS}. 
However, these approaches relied on implementation knowledge; they could not be applied to an arbitrary web service.
Prior researchers have not studied whether attackers can identify \REDOS vulnerabilities in a black-box manner.
If so, the engineering community should prioritize adopting \REDOS mitigations~\cite{Saarikivi2019SymbolicRegexMatcher,turovnova2020regex,Davis2021SelectiveMemo}.

In this paper, we describe the first black-box measurement methodology for \REDOS vulnerabilities (\cref{section:Methodology}).
We exploit software engineering practice, examining a previously-unstudied source of \REDOS information: the regexes that web services provide for use in client-side sanitization.
In the first step of our method, we collect the regexes provided in HTML forms (sampling popular websites) and in APIs (using a directory of services with OpenAPI specifications). 
Then, we analyze them locally for problematic worst-case behavior in a typical unsafe regex engine.
Finally, we ethically probe web services for \REDOS vulnerabilities.

\begin{figure*}[th!]
  \centering
  \begin{subfigure}{0.49\textwidth}
  \centering
  \includegraphics[width=0.80\columnwidth]{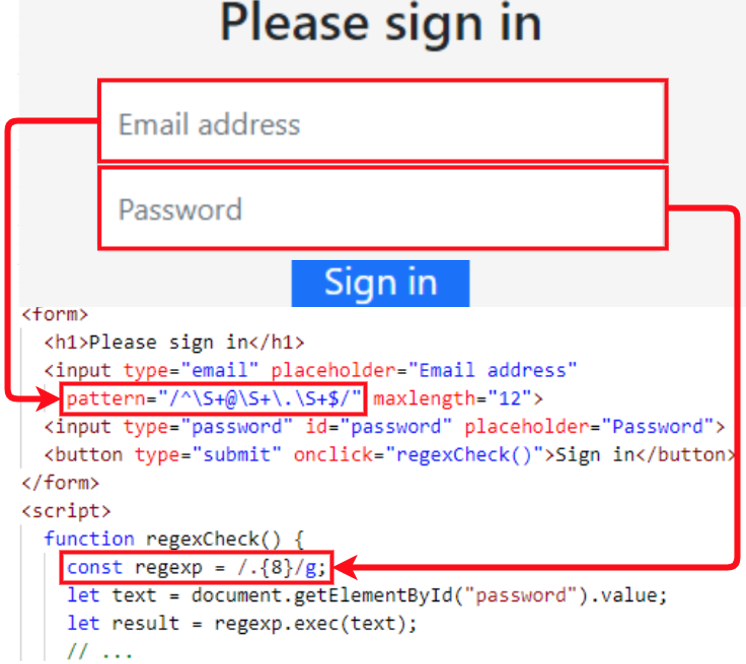}
  \caption{
    Sanitization in HTML form.
    Regexes can be applied using an input field's \code{pattern} attribute, and via JavaScript event handlers.
  }
  \label{figure:ClientSideSanitization-HTMLForm}
  \end{subfigure}%
  \hfill
  \begin{subfigure}{0.49\textwidth}
  \centering
  \includegraphics[width=.75\columnwidth]{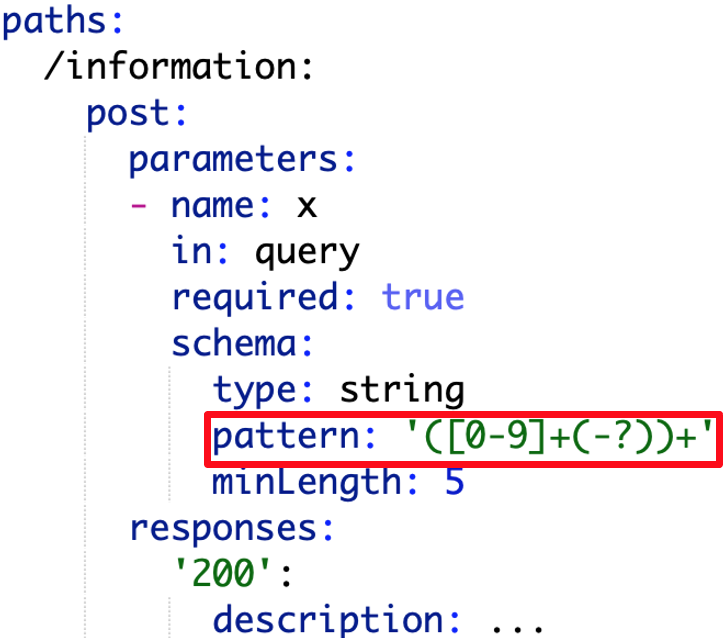}
  \caption{
    Sanitization in OpenAPI-based web API.
    Regexes are supported to encode string constraints in popular web API schema definition languages including OpenAPI~\cite{openapi}, RAML~\cite{raml}, and API Blueprint~\cite{api-blueprint}.
  }
  \label{figure:ClientSideSanitization-API}
  \end{subfigure}%
  \caption{
    The use of regexes for client-side sanitization in a web form and a web API.
  }
\label{fig:ClientSideSanitization}
\end{figure*}

Our findings indicate that emerging software engineering practices on the web expose web service providers to \REDOS (\cref{section:Results}).
Based on a sample of $N=1000$ popular websites, we report that web service providers do not reveal \REDOS vulnerabilities through their traditional HTML forms.
In contrast, web service providers reveal \REDOS vulnerabilities through API specification documents.
In our study of a live web services with OpenAPI specifications:
  there were 475 web domains;
  83 of them document their input sanitization regexes;
  and 6 of these (15 distinct subdomains) are vulnerable to \REDOS.
Web service providers publish much more information about their sanitization practices in their API specifications than in their traditional HTML forms.
To summarize our contributions:

\begin{itemize}
  \item We use the Consistent Sanitization Assumption to design the first black-box \REDOS measurement scheme for web services (\cref{section:Methodology}).
  \item We identify \REDOS vulnerabilities in several live web services (\cref{section:Results}). Comparing traditional HTML forms with the emerging approach of API specification, we report that current API specification practices expose web service providers to \REDOS.
  \item We describe the responses of engineering practitioners to the vulnerabilities we identified (\cref{section:Results-RQ4}). We contribute to the engineering practitioner community through a pull request to a major web API input sanitization library. The pull request has been merged and released. \footnote{See \url{https://github.com/ajv-validator/ajv/pull/1684} and \url{https://github.com/ajv-validator/ajv/pull/1828}.}
  \item We share a dataset of web sanitization regexes to complement existing regex datasets mined from other sources.
\end{itemize}

\myparagraph{\ul{Significance}:}
Three aspects of our research contributions are significant.
First, we establish the first black-box measurement methodology for \REDOS vulnerabilities in live web services.
Second, we use this methodology to identify insecure software engineering practices that affect a growing area of the web: APIs.
Third, we offer an anti-\REDOS patch that will benefit millions of dependent modules in the OpenAPI ecosystem.
Measurements drive change.

\section{Background} \label{section:Background}

\subsection{Web Services and Web Interfaces}
A web service is a software component (server) with which a user (client) can communicate over the Internet via a uniform resource identifier~\cite{FerrisWebServices}.
Common examples include the web services offered by YouTube and Amazon.
Clients interact with a web service using its interface, which is commonly defined in two ways:
  (1) a browser-based interface;
  and
  (2) an application programming interface (API). 

Most web services offer browser-based interfaces within their websites.
Websites are built using technologies such as HTML, CSS, and JavaScript, and displayed to clients through a web browser. To make websites responsive, web engineers provide interfaces, such as search boxes and login forms, to websites' users. Those interfaces allow users to send data to websites' servers to process the data and respond if needed. 
To build such user interfaces, engineers often use HTML forms~\cite{HTMLForm}, as depicted in~\cref{figure:ClientSideSanitization-HTMLForm}.


While browser-based interfaces target the general public, some web services support an Application Programming Interface (API) for automated interactions.
APIs are software interfaces that describe how different pieces of software should communicate with each other.
Through an API, a web service provider can give a more formal description of how to interact with the service.
This description enables engineers to develop software that interacts with the service programmatically.
APIs can be described with an informal text-based document, or with a schema definition language such as OpenAPI~\cite{openapi}, RAML~\cite{raml}, or API Blueprint~\cite{api-blueprint} (cf.~\cref{figure:ClientSideSanitization-API}).
API semantics may be explicit or implied, \eg using the conventional meaning of REST verbs to develop a REST-ful API~\cite{Fielding2000REST}.

By their nature, web services interact with untrusted clients.
It is therefore standard engineering practice to sanitize any client input, whether it comes via a browser-based interface or an API~\cite{SomeSoKAndCSURAboutInputSanitization}.
Since the client controls this input, sanitization ought always to be performed as part of the server's logic.
However, some web services also publish input sanitization logic to their clients, to reduce network traffic and give clients feedback about invalid requests.

\cref{fig:ClientSideSanitization} depicts common forms of sanitization published to clients.
The HTML form in~\cref{figure:ClientSideSanitization-HTMLForm} illustrates the two ways to perform client-side sanitization for HTML forms: HTML-based and JavaScript-based~\cite{Client-side-form-validation}.
Using HTML-based form validation, engineers can enforce attributes on various HTML tags in HTML forms.
The attribute of interest in this work is the ``pattern'' attribute, which
lets an engineer specify the language of legitimate input. 
For more sophisticated checks, JavaScript-based validation supports custom client-side validation logic applied on a relevant event such as an attempted form submission. Just like HTML-based form validation, regexes can also be used in JavaScript-based validation to check the validity of an input string.
Meanwhile, \cref{figure:ClientSideSanitization-API} depicts an OpenAPI-style API definition.
Similar to HTML forms, API schema documents may constrain request headers, payload structure, and field types and valid values.
These constraints may indicate enumerations, numeric ranges, string lengths, and --- of interest in our study --- regexes prescribing string input languages.

Client-side sanitization can help legitimate users debug their requests, \eg via feedback from the web browser or an automatically generated client API driver.
However, malicious clients can bypass client-side sanitization and send unsanitized content to the web service, so services must sanitize again on the server side~\cite{Offutt2004BypassTesting}.

\subsection{Regexes and Regex-based Denial of Service} \label{subsection:Regexes and Regex-based Denial of Service}

Our work measures a form of \emph{denial of service} that web services risk as a result of sharing input sanitization regexes with their clients.

\myparagraph{Denial of Service Attacks:}
A denial of service attack consumes the resources of a service so that legitimate access to the service is delayed or prevented~\cite{nieles2017}. 
There are many types of denial of service attacks, varying in the resource exhausted and the exhaustion mechanism. 
For example, attacks might exhaust network resources (\eg distributed denial of service) or computational resources (\eg algorithmic complexity attacks~\cite{Crosby2003AlgorithmicComplexityAttacks}).

Regex-based denial of service (\REDOS) is a denial of service attack that exhausts computational resources by exploiting the worst-case time complexity of an unsafe regex engine.

\myparagraph{Unsafe Regex Engines:}
A regex describes a language (a set of strings)~\cite{Hopcroft2006AutomataTextbook}.
To determine whether a string matches a regex, membership testing is conducted by a system component called a regex engine~\cite{Friedl2002MasteringRegexes}. 
Most programming languages embed a custom regex engine for efficient interactions with the programming language's string encoding.
These regex engines support diverse features with complex semantics~\cite{Goyvaerts2016LanguagesWithRegexSupport}.
To reduce implementation and maintenance costs, some regex engine developers chose designs that favor simplicity over safety~\cite{Davis2021SelectiveMemo} --- they use a predictive parsing algorithm with backtracking~\cite{aho2013compilers,Cox2007RegexMatchingCanBe,Davis2019LinguaFranca}.
The emphasis on simplicity comes at a cost: this algorithm has high time complexity, polynomial or exponential in the worst case.

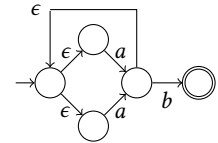
\begin{figure}[ht] 
    \centering 
      \centering
      \tikzset{invisible/.style={minimum width=0mm,inner sep=0mm,outer sep=0mm}}
      \begin{tikzpicture}
        [initial text=]
        \tikzstyle{every state}=[inner sep=0pt, minimum size=0.4cm]


        \node[state, initial, initial distance=0.25cm] (q1) {$$};
        \node[state, above right = 0.4cm of q1] (q2) {$$};
        \node[state, below right = 0.4cm of q1] (q3) {$$};
        \node[state, below right = 0.4cm of q2 ] (q4) {$$};
        \node[state, right = 0.4cm of q4, accepting ] (q5) {$$};

        \draw[->] (q1) edge [] node [above, pos=0.25] {\tt $\epsilon$} (q2);
        \draw[->] (q1) edge [] node [below, pos=0.2]  {\tt $\epsilon$} (q3);
        \draw[->] (q2) edge [] node [above, pos=0.75]  {\tt $a$} (q4);
        \draw[->] (q3) edge [] node [below, pos=0.75]  {\tt $a$} (q4);
        \draw[->] (q4) edge [] node [below, pos=0.5]  {\tt $b$} (q5);

        \node[label=left:{$\epsilon$}, above= 0.75cm of q1,invisible] (above1) {};
        \draw (q4) |- (above1);
        \draw[->] (above1) -- (q1);
      \end{tikzpicture} 
    \caption{
        This non-deterministic finite automaton (NFA) corresponds to the regex 
        \emph{\texttt{(a|a)*b}}.
        Using the typical Spencer algorithm~\cite{Spencer1994RegexEngine}, viz. a predictive parse with backtracking, the search space can be exponential in the length of the input.
        For example, consider the behavior on input ``$a$...$a$'' (length $k$).
        The regex does not match this string; the backtracking algorithm explores all $2^{k}$ failing paths.
    }
    \label{figure:ExponentialRegex} 
\end{figure}

The high time complexity of the standard regex engine algorithm is triggered by a problematic combination of a regex and an input string.
These regexes are \emph{super-linear}; there are input strings $w$ that incur time complexity super-linear in the length of $w$.
Most regexes in this class are \emph{problematically ambiguous}~\cite{Allauzen2008TestingAmbiguity,Rathnayake2014rxxr2,Weideman2016REDOSAmbiguity,Wustholz2017Rexploiter}.
Because they are ambiguous, these regexes can match a string in multiple ways.
The typical regex engine's backtracking search algorithm will explore all potential matching paths before returning a mismatch.
When the number of paths or the cost of each path depends on the length $\vert w \vert$, the result can be super-linear time complexity. 
\cref{figure:ExponentialRegex} illustrates an example of exponential behavior.
Many researchers have proposed tools to identify regex-input pairs with worst-case polynomial or exponential behavior~\cite{Brabrand2010DisambiguatingRegexes,Berglund2014REDOSTheory,Rathnayake2014rxxr2,Sugiyama2014RegexLinearityAnalysis,Weideman2016REDOSAmbiguity,Wustholz2017Rexploiter,Petsios2017SlowFuzz,Sulzmann2017DerivAmbig,Shen2018ReScueGeneticRegexChecker,liuRevealerDetectingExploiting2021}.

\myparagraph{Regex-Based Denial of Service:}
Regex-based denial of service (\REDOS), exhausts computational resources by exploiting the algorithmic complexity of regex engines~\cite{,Crosby2003REDOS}.
When client-controlled input can trigger worst-case regex behavior (\cref{fig:AttackOverview}), it can be harmful.
For example, in 2016 Stack Overflow had a system-wide outage due to \REDOS~\cite{StackOverflow2016REDOSPostMortem}, and in 2019 Cloudflare had a \REDOS outage that affected thousands of its customers~\cite{Cloudflare2019REDOSPostMortem}.
While \REDOS vulnerabilities directly impact compute resources, depending on the system design they may impact higher-order resources.
For example, many web services multiplex between clients, \eg event handler threads~\cite{Davis2018NodeCure}, and in these designs a \REDOS attack will be more effective. 
However, even with near-perfect client isolation, \eg via AWS Lambda, algorithmic complexity attacks like \REDOS provide attackers with an asymmetric attack/defense cost ratio to inflict economic damage~\cite{sides2015yo,somani2017combating}.

Slow reachable server-side regexes are a security risk.
More formally~\cite{davis2020impact}, a \REDOS attack requires four \emph{\REDOS Conditions} of a victim web service:
(1)~It accepts \emph{attacker-controlled input};
(2)~It uses a server-side \emph{super-linear regex} on this input;
(3)~It uses an \emph{unsafe regex engine};
and
(4)~It has \emph{insufficient mitigations} to insulate other clients from slow server-side regex matches (\eg timeouts).
Although mitigations may reduce the service's \REDOS risk, they do not eliminate it~\cite{sides2015yo,somani2017combating}.
Therefore, if a service meets Conditions 1--3, we consider it vulnerable to \REDOS.

For example, suppose a web service is built with the Node.js framework.
If it has a reachable super-linear regex, then Conditions 1 and 2 are met.
The regex may be evaluated on Node.js's unsafe default regex engine (Condition 3).
Its slow performance would then affect other clients due to client multiplexing on the Node.js event loop (Condition 4)~\cite{Ojamaa2012NodeJSSecurity,Davis2018NodeCure,Staicu2018REDOS}.

\subsection{Prior Empirical Studies on \REDOS}
The two previous empirical measurements of the extent of \REDOS in practice have used a strong threat model: that the attacker controls the input and also has server-side implementation knowledge.
Under that model, \Wustholz \etal identified \REDOS vulnerabilities in open-source Java applications using program reachability analysis, reporting that many Java applications used reachable super-linear regexes~\cite{Wustholz2017Rexploiter}.
Staicu \& Pradel exploited knowledge of the open-source JavaScript software supply chain~\cite{Raymond2000,Wittern2016NPM} to predict \REDOS vulnerabilities in web services that use Express, the Node.js server-side framework --- 10\% of the services they probed had \REDOS vulnerabilities~\cite{Staicu2018REDOS}.

Although these studies document the risks of \REDOS in practice, their methodologies depend on knowledge of web service internals, and are unsuitable to larger-scale probing of web services in a black-box manner.
Finally, studies by Davis \etal measured the extent of super-linear regexes within the software module supply chain~\cite{Davis2018EcosystemREDOS,Davis2019LinguaFranca,Davis2019RegexGeneralizability}, and do not shed light on web service vulnerabilities.

%


\section{Attack and Research Questions} \label{section:RQs}

\myparagraph{Threat Model:}
Our primary interest is to answer the question: \emph{To what extent do \REDOS vulnerabilities exist in live web services?}
As discussed in~\cref{section:Background}, prior work has measured the possibility of \REDOS (through module analysis) and the presence of \REDOS (through white-box analysis).
Thus far we lack a methodology to measure the risk that \REDOS poses to general black-box web services.

We therefore assume the weakest reasonable threat model.
First, we suppose the attacker controls only the \emph{input} (Condition 1).
Second, we suppose the attacker does \emph{not} have access to the web service's server-side logic.
Under this threat model, the primary difficulty is in identifying a reachable super-linear regex to satisfy Condition 2.\footnote{In the 2000s there were many CVEs for web services that allowed users to specify a regex to be evaluated on the server side (Condition 2), and they used an unsafe regex engine. Such CVEs are now rare, so we suppose a weaker threat model.}
Once such a regex is identified, the attacker can tailor their input to the regex, then use probes to experimentally determine whether Conditions 3 and 4 are met.

When engineering teams evaluate their own services, this threat model may be needlessly restrictive.
But it may imitate the perspective of engineers assessing the risks of incorporating a third-party service or component into their product, or that of adversaries, penetration testers, and ``security-scanning-as-a-service'' vendors.

\begin{figure}[h!]
    \begin{center}
    \includegraphics[width=0.75\columnwidth]{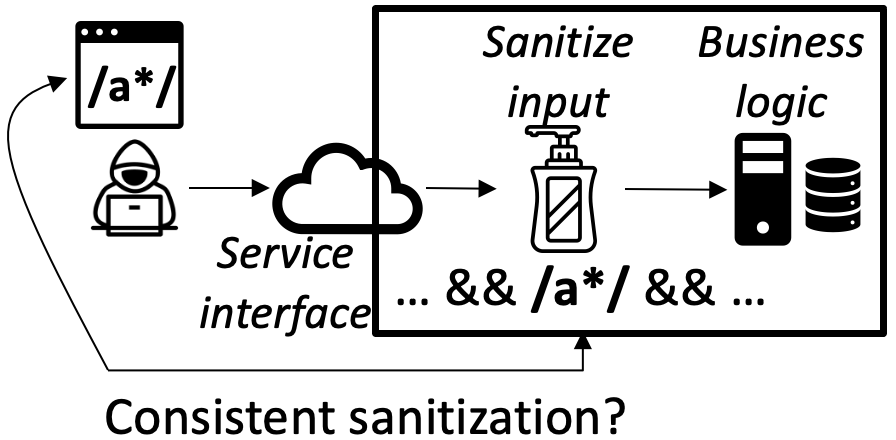}
    \end{center}
    \caption{
      Web service model.
      For \REDOS,
        the attacker identifies a web service with a vulnerable sanitization regex,
        then transmits data that
          (a) passes preceding constraints, then
          (b) triggers the worst-case behavior of the regex engine.
    }
    \label{fig:AttackOverview}
    \vspace{-0.1cm}
\end{figure}

\myparagraph{Sanitization-Based \REDOS Attacks:}
Given this constraint, we propose \emph{sanitization-based \REDOS attacks}.
As noted in~\cref{section:Background}, while web services do not typically publish their server-side implementations, some of them do publish client-side input sanitization logic.
We adopt the \emph{Consistent Sanitization Assumption} (\cref{fig:AttackOverview}): following engineering conventions, the client-side sanitization logic that a web service publishes is a subset of its server-side sanitization logic.
This assumption implies that a super-linear regex used in client-side sanitization logic will fulfill \REDOS conditions 1 and 2: this regex will be applied to attacker-controlled input on the server-side.
If true, \REDOS vulnerabilities can be discovered by finding super-linear sanitization regexes in client-side sanitization logic, and then probing web services to test the remaining conditions.

\myparagraph{Research Questions:}
We conduct the first black-box web measurement study of the extent of \REDOS-vulnerable web services in practice.
Our operationalized research questions are:

\begin{RQList}
    \item[RQ1:] How common is regex-based client-side input sanitization? (\REDOS Condition 1)
    \item[RQ2:] What proportion of these regexes would be super-linear in an unsafe regex engine? (\REDOS Condition 2)
    \item[RQ3:] To what extent do these regexes exhibit super-linear behavior in live web services? (\REDOS Condition 3)
    \item[RQ4:] How does the web service community mitigate these problematic super-linear regexes? (\REDOS Condition 4)
\end{RQList}

Research ethics constrain us from fully characterizing the extent of \REDOS vulnerabilities with this method.
Using a black-box approach, we can measure whether a service meets the first three \REDOS Conditions:
  (1) it evaluates untrusted input,
  (2) on a problematic regex,
  (3) with a slow regex engine.
However, we cannot comment on (4) the mitigations the service has in place, whether architectural or runtime.
Web services are opaque.
Their mitigations are difficult to assess without launching a denial of service attack.
In addition, because our method relies only on publicly-accessible service information, we cannot comment on the existence of hidden \REDOS vulnerabilities such as those identified by Staicu \& Pradel via implementation inference~\cite{Staicu2018REDOS}.
With these caveats, our method lets us measure black-box web services for a new perspective on the risks of \REDOS in the wild.


\section{Methodology} \label{section:Methodology}

\begin{figure}[h]
    \centering
    \includegraphics[width=0.95\columnwidth]{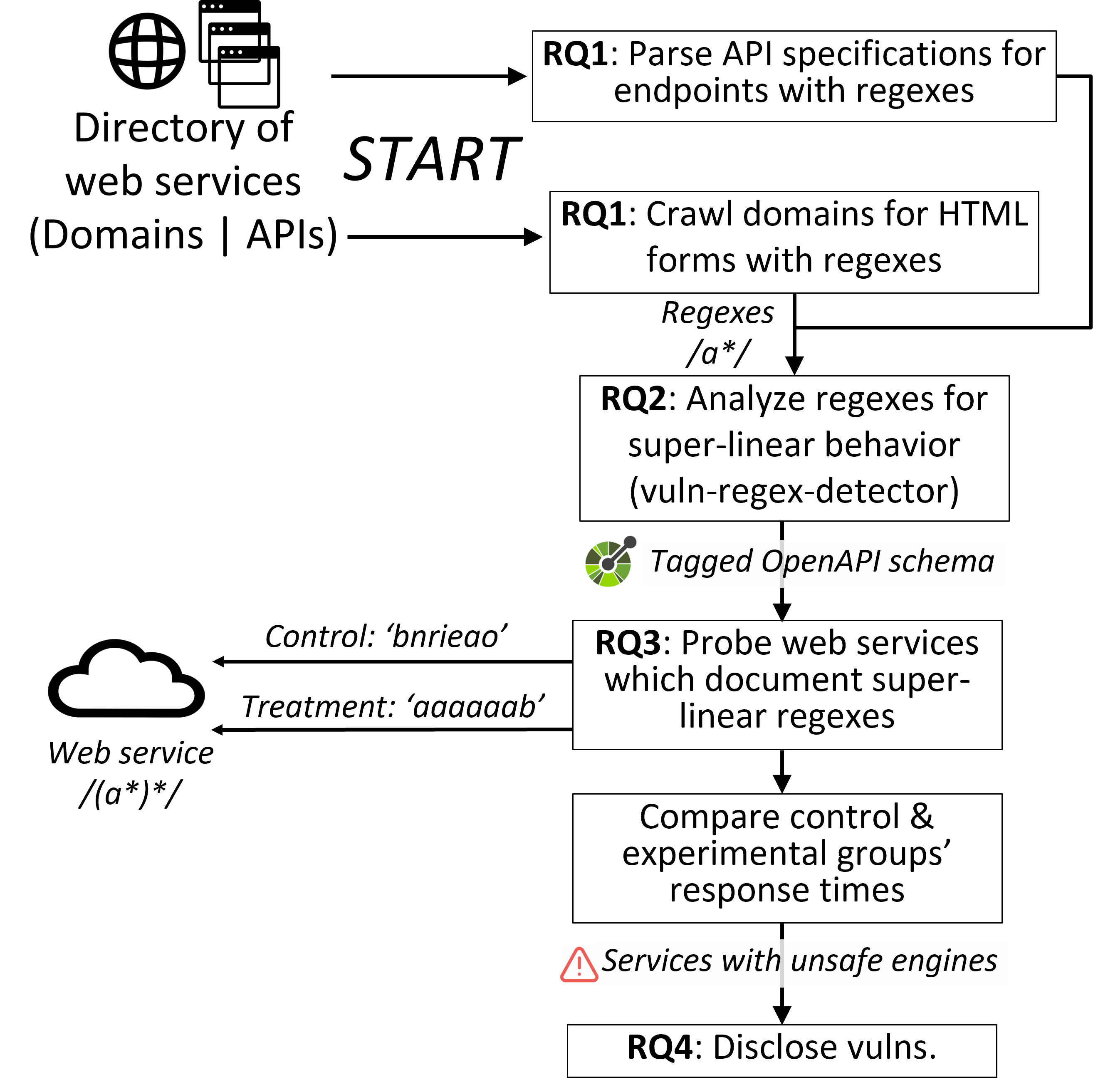}
    \caption{
        Overview of the study design.
    }
    \label{fig:ResearchMethodologyOverview}
\end{figure}

Our methodology is shown in~\cref{fig:ResearchMethodologyOverview}.
Given a web interface, first we analyze its client-side input sanitization logic to determine the input fields and the regexes applied to them (\cref{sec:Methodology-RQ1}).
Next, we identify any super-linear regexes (\cref{sec:Methodology-RQ2}). 
Then, we ethically probe the web service for \REDOS vulnerabilities (\cref{sec:Methodology-RQ3}).
Finally, we interact with web service engineers to understand their perspectives (\cref{sec:Methodology-RQ4}).

\subsection{Web Service Selection} \label{sec:Methodology-ServiceSelection}

As discussed in~\cref{section:Background}, web services commonly offer two kinds of interfaces: HTML forms and APIs.
We measure \REDOS vulnerabilities through both kinds of interfaces.

\myparagraph{HTML Form Interfaces:}
For HTML form interfaces, we examine the forms within a random sample of \WebFormsSampleSize domains from the top \WebFormsTopXPopulationSize domains according to the Tranco Top 1M ranking~\cite{pochattranco2019}.
The Tranco directory is a website popularity ranking designed to address shortcomings of the Alexa list~\cite{AlexaTopSites}.


\myparagraph{APIs:}
API-based interfaces are less common than HTML form interfaces.
To obtain sufficient data, we focused on the popular OpenAPI~\cite{openapi} schema description language for REST-ful~\cite{Fielding2000REST} APIs.
We determined popularity using web searches and GitHub stars~\cite{Borges2018GitHubStars}; OpenAPI has 23K stargazers, about three times more than the next-most-common API language, API Blueprint~\cite{api-blueprint}.

Following prior work on specification measurement~\cite{Wittern2019EmpiricalGraphQL}, we used the \code{apis.guru} directory~\cite{apisguru} to obtain OpenAPI documents.


\subsection{RQ1: Input Sanitization Regexes} \label{sec:Methodology-RQ1}

In this part of the study, our goal is to measure how frequently web services include regexes as part of their client-side input sanitization.
To do this, we build a list of live web services and the regexes they use in their client-visible input sanitization.
As depicted in~\cref{fig:ClientSideSanitization}, software engineers can impose similar input sanitization using the two interface types.
The mechanism for specifying this sanitization differs by interface type.

\myparagraph{HTML Form Interfaces:}
For HTML forms, regexes may occur in two places (\cref{table:HTMLClientSideInputSanitization}).
First, in the relevant form field, the string language of valid input can be described by a JavaScript-dialect regex using the HTML5 \code{pattern} attribute.
Second, JavaScript logic can be applied to form fields, \eg on data entry or button press events.
This logic can impose constraints including regex tests.

\begin{table}[ht]
\centering
\caption{
    Regex-based sanitization in browser interfaces~\cite{MDNDocs}.
    \cref{figure:ClientSideSanitization-HTMLForm} also shows both types of HTML Form regexes.
}
\label{table:HTMLClientSideInputSanitization}
\noindent\begin{tabular}{cc}
\toprule
\textbf{Type}       & \textbf{Functions} \\
\midrule
HTML attribute           & \code{pattern} attribute \\
\midrule
JS: String          & String.\{match, matchAll, search\} \\
JS: RegExp          & RegExp.\{test, exec\} \\
\bottomrule
\end{tabular}
\end{table}

The HTML forms that comprise a web service's browser interface may occur on any page.
We used the Apify web crawler~\cite{Web-Scraping-Apify} to crawl each website from its homepage and identify forms.
To balance our desire for detailed crawls with the need to not take resources from real users, we used a maximum crawl depth of \ApifyMaxCrawlDepth and fully crawled \ApifyPercentFullyCrawled of the crawled web sites.

After identifying each form, we determine the regexes it uses in its client-side sanitization.
We define this set as: ``\emph{any regex that is applied to any form field prior to sending form content to the server}''.
We statically extract the regexes given as form attributes.
We use a simple dynamic taint analysis to identify regex constraints in JavaScript logic.
First, we monkey-patch the client-side JavaScript regex functions (\cref{table:HTMLClientSideInputSanitization}) to log each regex-string pair. This is done by modifying those functions' definitions in the browser so that, each time the functions are called, we have access to their input parameters.
Then, we drive a web browser via OpenWPM~\cite{OpenWPM-2021}, which is a software that can control browsers programmatically, to populate form fields with unique values and simulate a button press. The forms and buttons are detected by parsing the HTML code of each webpage. We use a proxy to discard the resulting form traffic so that we do not spam the web service. 
Then, inspecting the monkey patch traces, we identify the regexes applied to each form field.
This may be a subset of the desired set --- our approach omits any regexes that are applied to the substrings of form fields, \eg logic that splits an email and checks a property on the username. 

Unsatisfactory form field values may lead the browser to reject our form \emph{before} our program instrumentation is triggered.
To reduce these cases, we solve the constraints encoded within HTML form attributes, \eg integer constraints directly and regex constraints using Z3~\cite{de2008z3}, although JavaScript-based constraints may still fail.

\myparagraph{APIs:}
For API-based interfaces we have the same goal: to identify the set of regexes that constrain client input.
For such an interface, client input can appear in HTTP headers, endpoints, query strings, and request bodies.
In typical API schema definition languages, including OpenAPI, engineers can set regex-based constraints on string inputs.
We parse a schema and identify the regex(es) that constrain any string inputs referred to by at least one request schema.


\subsection{RQ2: Super-Linear Sanitization Regexes} \label{sec:Methodology-RQ2}

Our next goal is to measure the proportion of client-side input sanitization regexes that present a \emph{potential} \REDOS vector.
We lack knowledge of a web service's server-side implementation, including its choice of regex engine.
A regex is a potential \REDOS vector if it has super-linear worst-case behavior in \emph{some} regex engine. 

To identify a super-linear regex, we apply the ensemble of state-of-the-art super-linear regex analyses supported by Davis \etals tool, \code{vuln-regex-detector}~\cite{Davis2019LinguaFranca,vuln-regex-detector}. 
These analyses vary in their 
soundness and completeness, so we dynamically test any potentially super-linear regex in a representative unsafe regex engine.
Davis \etal found the Java, JavaScript, and Python regex engines were in the most unsafe class of engines~\cite{Davis2019LinguaFranca}.
Although the Java~\cite{van2021memoized} and JavaScript-V8~\cite{V8UpdatebidlingmaierAdditionalNonbacktrackingRegExp2021} regex engines have recently been optimized, the Python regex engine has not.
We therefore tested regexes in the Python regex engine (Python v3.8.10).

We define a regex as super-linear using the definition from algorithmic complexity theory~\cite{cormen2009introduction}: when it exhibits a more-than-linear increase in match time in the input length $\vert w \vert$ as we increase the number of ``pumps'' of the attack input strings. 
We further distinguished the degree as ``high-complexity'' and ``low-complexity'' depending on the number of pumps necessary to yield substantial matching times, similar to Davis \etal~\cite{Davis2019LinguaFranca}.

\subsection{RQ3: Use of Unsafe Regex Engine} \label{sec:Methodology-RQ3}

By now we have identified interfaces in live web services that publish a super-linear regex used on client input.
Our next goal is to understand the proportion of these regexes that are \emph{actual} \REDOS vectors, \ie testing whether the server uses these regexes in an unsafe regex engine.

\subsubsection{Measurement algorithm}

\begin{figure}
    \begin{center}
    \includegraphics[width=0.85\columnwidth]{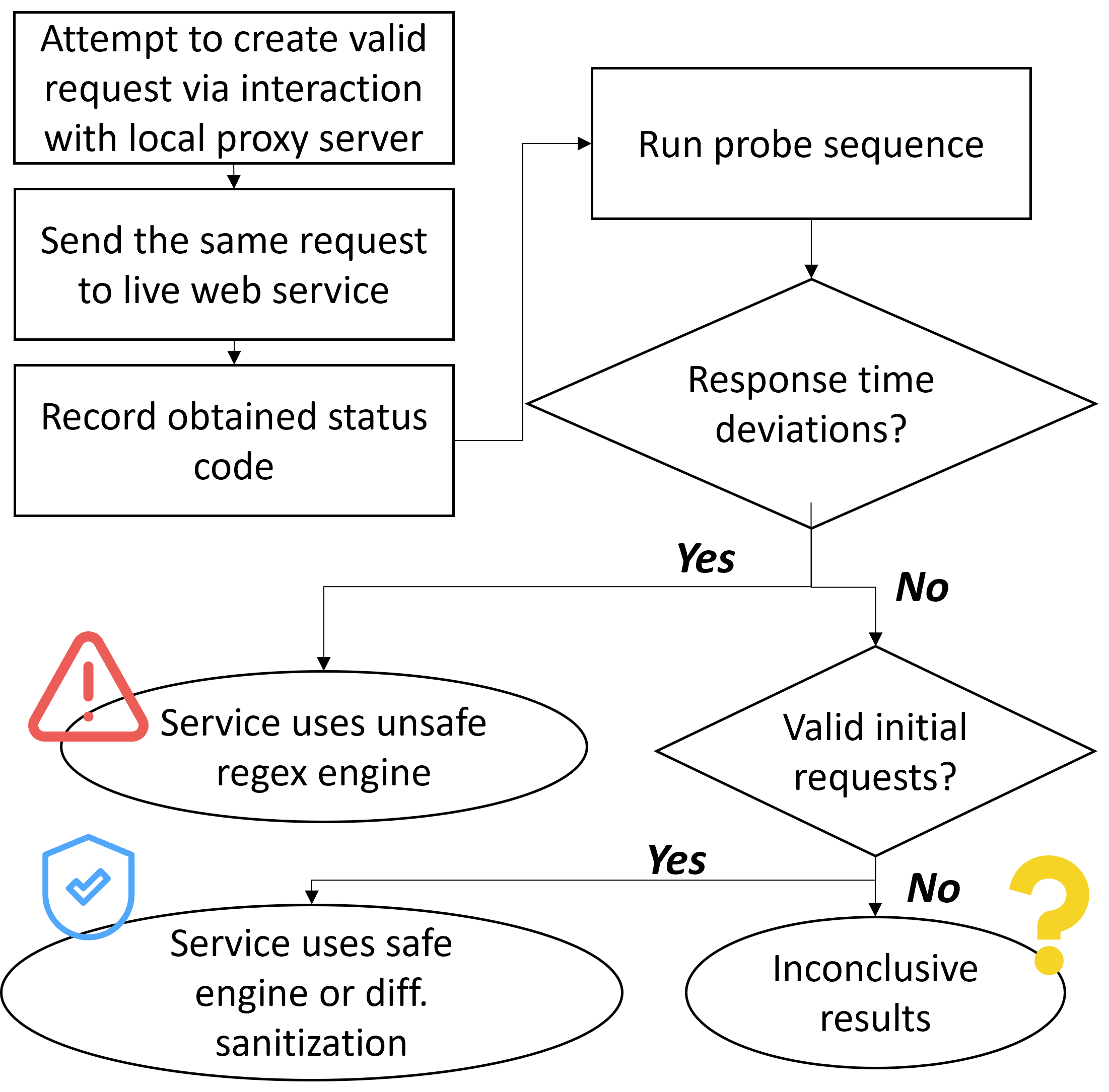}
    \end{center}
    \caption{
        Measurement process for \REDOS Condition 3.
        We begin with a web service interface, a client input field, and a super-linear regex applied to that input on the client side.
        As a reachability check, we seek a valid status code from a proxy and then from the live web service.
        After a probe sequence of treatment and control requests, 
        a decision tree follows.
    }
    \label{fig:oracle-flowchart}
\end{figure}

As depicted in~\cref{fig:oracle-flowchart}, our measurement algorithm attempts to
  (1) reach the relevant logic in the interface's server-side implementation; and then
  (2) identify linear vs. super-linear regex behavior on the server-side
  while
  (3) avoiding actually conducting a denial of service attack.

We assume that these live web services follow standard HTTP semantics~\cite{Fielding2000REST}, in particular that if a web service responds to a client request with a success return code (\code{2XX}) then the request was legitimate.
We assume such a request has passed all server-side sanitization.
If the Consistent Sanitization Assumption holds, then this means that any client-side regexes were also applied to the relevant input field(s) on the server side.
Requests can then be sent to determine whether these regexes exhibit super-linear behavior.

It is possible that the target regex can be reached even without a successful baseline request, but it depends on the cause of the failure.
For example, the target regex constraint might be applied \emph{before} the failing condition.
Thus, if no successful request can be crafted, super-linear behavior may still be observed; but if we observe linear-time behavior, the results are inconclusive. 
If we cannot identify a valid request, we use an invalid one.

\subsubsection{Crafting a valid client request:}

\myparagraph{HTML Form Interfaces:}
For HTML form interfaces, we identified and satisfied the constraints embedded in HTML form attributes as part of RQ1 (\cref{sec:Methodology-RQ1}).
To test constraint validity for these interfaces, if a request reaches our HTTP proxy, then we conclude that it passed the client-side sanitization.
We send requests to the web service using the Python \code{requests}~\cite{requests} module. 

\myparagraph{APIs:}
For APIs, the first problem is reaching the target endpoint.
For example, reaching an endpoint like \code{/home/\{USER\_ID\}/photos} requires dynamic information (a \code{USER\_ID}) obtained from the web service.
After this, we must satisfy the constraints associated with the fields of the endpoint in question --- among other limitations, RESTler's fuzzing strategy may not satisfy the regex constraint that is present for these endpoints.
For reachability, we implemented our API analysis as a ``checker'' plug-in within Microsoft's state-of-the-art REST API fuzzer, RESTler, version v7.4.0~\cite{Atlidakis2019RESTler}.
RESTler uses API conventions to determine dependency relationships between endpoints, with a simple fuzzing dictionary to attempt to satisfy each endpoint's constraints.
Once RESTler reaches the target endpoint, our plug-in is called to populate the request fields.
We use \code{json-schema-faker}~\cite{json-schema-faker} for this purpose, and then populate the \REDOS-relevant field in the endpoint of interest.

To test constraint satisfaction for APIs using RESTler with our plug-in, we use the Prism mock server tool v4.2.6~\cite{Prism} to validate requests and generate mock responses according to the OpenAPI specification of interest.
Prism returns codes in the \code{4XX} range if any constraints are missing.
We treat other codes as an indicator of satisfied constraints.



\subsubsection{Ethical \REDOS probing}

We send probe requests by injecting \emph{probe strings} into a previously-sent valid request, or an invalid request if no valid status codes were obtained during the probing experiment.
These probe strings are assembled from templates produced by the regex analysis component.
The templates contain three strings: a prefix, suffix, and a pump.
A probe string is a concatenation of the prefix, one or more repetitions of the pump, and the suffix, and triggers the worst-case behavior of a regex during a mismatch.
Each additional pump increases the match time super-linearly in an unsafe regex engine.

We devise a five-stage probing experiment based on that of Staicu \& Pradel~\cite{Staicu2018REDOS}.
Our overall goal is to identify ``treatment'' input strings that yield a $\geq1$ second increase in response time relative to a comparable ``control'' string --- without causing substantial slowdowns for normal clients.
To that end:
(1) We identify an initial set of treatment input strings with a range of matching times (200ms to 3s) using the performance of the ``maximally unsafe'' Python regex engine on our workstation.
These input strings should yield a small but measurable time difference in an unsafe regex engine.
(2) We send \ProbeNWarmupRequests (preferably valid) warm-up requests to address response time noise caused by first-time operations, such as cache filling. 
These use a valid request if we identified one, else an invalid one.
(3) For each timing configuration, we send an \emph{experiment} sequence of \ProbeNTreatmentAndControlRequests requests with the vulnerable field populated with a probe string, and a \emph{control} sequence of \ProbeNTreatmentAndControlRequests requests with that field populated with randomly generated strings of the same lengths (these run in linear time).
(4) Both groups of requests are expected to fail at the same stage of validation, viz. the regex constraint.
If the median round-trip response time in the treatment group is substantially larger than in the control group, we conclude that the web service being probed uses an unsafe regex engine.
Specifically, we look for a 1-second increase in the median round-trip time for the treatment group.
(5) If the service is using an unsafe regex engine, its server hardware or runtime timeouts may affect the actual response time relative to our predication.
If we observe a \code{5XX} response code or a response time greater than 5 seconds, we halt the experiment to avoid harm and consider the regex engine unsafe. 
Conversely, if the treatment group exhibits deviations but below the 1-second threshold, we manually explore a longer probe sequence.

We identified three known mitigations that can mask unsafe regex engine behavior under this protocol.
First, server-side rate limiting could delay our probes regardless of their content.
  Although rate limiting would presumably not cause the treatment-control \emph{deviations} that we measure, we sent no more than 1 request per second to account for this possibility.
Second, caching --- either of the validation outcome, or of end-to-end results --- could cause only the initial query at each probe size to be slow.
  We manually observed one case of this form.
Third, a recent approach can identify the signatures of anomalously slow regex input~\cite{baiRuntimeRecoveryWeb2021},
  although we are not aware of applications of this technique in practice.

\subsection{RQ4: \REDOS Mitigation} \label{sec:Methodology-RQ4}

In this part of the study, our goal was to understand the perspective of the web service engineering community on the use of super-linear regexes in server-side input validation.
We assessed this
  constructively (proposed mitigation),
  as well as
  in a responsibly destructive manner (vulnerability identification and disclosure).

Constructively, we assessed the state of \REDOS mitigations in OpenAPI-based automatic client sanitization libraries.
We found an absence of mitigations, proposed one, and report on our findings.

``Destructively'', we contacted the owners of live web services for which \REDOS Conditions 1--3 held: cases where our experiments identified super-linear regex performance in live web services.
Since \REDOS is a security problem, we disclosed such issues to web service engineers using their documented route, \eg the \code{security@domain.com} email for major companies.
We informed them of a super-linear regex in their client-side input sanitization, presented the attack format, and gave a minimal example.
We asked whether they considered this a security vulnerability in their service, and what mitigations they had in place.

\subsection{Automating RQ1--RQ3}

We automated most parts of this measurement process, using existing tools as indicated.
We manually intervened when this automation failed.
This was particularly notable for the APIs; these services vary in the accuracy of the semantics that they encode in the OpenAPI schema.
We intervened to repair schema syntax, authenticate, and supply values RESTler could not obtain (\eg some resource ID values or under-documented constraints).
Some interventions were guided by a service's error messages.

For one web service with a particularly complex API, we used the official client SDK, documentation, and browser interface to craft valid requests to endpoints with super-linear regexes.
\section{Results and Analysis} \label{section:Results}



For security measurement purposes, we are interested in understanding the extent to which a given \emph{web service} is potentially vulnerable to \REDOS attacks.
The attack surfaces in question are clear --- individual HTML forms and API endpoints.
However, services may employ the same sanitization policy across multiple surfaces.
We present results aggregated by web domain, as well as aggregated by subdomains where appropriate.

\subsection{RQ1: Published Sanitization Information} \label{section:Results-RQ1}

\FindingsBox{1}{
  Web services frequently use regexes to sanitize input on the client side.
  \WebFormRegexUsingDomainCount of the 696 reachable HTML form domains do so,
   as do
  \APIRegexUsingDomainCount of the \ApisGuruDomainCount studied API domains.
}

\begin{table}[ht]
\centering
\caption{
    Use of regexes in client-side input sanitization.
    \emph{Domains and sub-entities}:
    For HTML forms, we report the number of web domains and web pages that apply client-side regexes to any form fields.
    For APIs, we report by domain and subdomain.
}
\label{tab:regex-usage}
\newlength\qq
\setlength\qq{\dimexpr 1\columnwidth -2\tabcolsep}
\noindent\begin{tabular}{cccc}
\toprule
\textbf{Interface type}  & \textbf{\# Domains} & \textbf{\# Sub-entities}         \\
\midrule
HTML form               & \WebFormRegexUsingDomainCount (\WebFormRegexUsingDomainPercent) & \WebFormRegexUsingPageCount (\WebFormRegexUsingPagePercent) \\
API                     & \APIRegexUsingDomainCount (\APIRegexUsingDomainPercent) & \APIRegexUsingSubdomainCount (\APIRegexUsingSubdomainPercent) \\
\bottomrule
\end{tabular}
\end{table}
\myparagraph{HTML Forms:}
 We crawled \WebFormsSampleSize domains sampled randomly from the Tranco Top 1M list.
 Through web crawling, we found at least one web page for \NumOfDomainAtLeastOnePage of those web sites.
 Our crawler failed on the remainder, \eg blocked by the service, and we omit them from the following statistics.
 Among the crawled \NumOfDomainAtLeastOnePage domains,
 the median number of pages per domain was \MediumNumPagePerDomain, and the median number of forms per domain was \MediumNumFormPerDomain.

\myparagraph{APIs:}
We obtained 2231 documents from apis.guru. 
\APIRegexValidatingDocuments documents contained at least one operation with a regex validation constraint.
These documents corresponded to \APIRegexUsingDomainCount web services with unique domain names, out of \ApisGuruDomainCount web services. 
The median number of documents per domain and per subdomain are 1.

\myparagraph{Analysis:}
We observed substantial variation in the number of \emph{unique} regexes amongst the input sanitization regexes.
For HTML form's pattern attribute regexes, there were \WebFormTotalPatternNumber total regex uses but only \WebFormUniquePatternNumber unique regexes.
Substantial regex re-use across websites is consistent with the findings of Hodov\'an \etal~\cite{Hodovan2010WebRegexes}, who examined the regexes parsed during browsing sessions.
This repetition may be the result of client-side library or framework re-use, with the regexes originating in web frameworks or JavaScript libraries rather than independently authored by many engineers.
In marked contrast to the duplication of regexes in HTML forms, in the API documents there were \APITotalRegexUseNumber total regex uses and \APIUniqueRegexNumber unique regexes.

For HTML forms, we also note that most regexes were employed in JavaScript logic which was used by \NumOfDomainUseRegexFromJS domains.
Only \NumOfDomainUsePattern of domains used the HTML5 \code{pattern} attribute in any form.
\pagebreak
\subsection{RQ2: Super-Linear Regexes} \label{section:Results-RQ2}

\FindingsBox{2}{
    Super-linear regex usage varies widely by interface type.
    Among the \NumOfDomainUseRegex domains with regexes in their HTML forms, only \DomainWithVulnRegex (\DomainWithVulnRegexWithinDomainWithRegex) use a super-linear regex.
    Meanwhile, among the \APIRegexUsingDomainCount regex-using API domains, \SLRegexDomainCountAPI domains (13\%) use a super-linear regex in at least one constraint.
}

\myparagraph{HTML Forms:}
We identified \VulnRegex super-linear regexes on \DomainWithVulnRegex distinct web services. Each vulnerable regex appears on exactly one service. Three web services had a super-linear regex on one page each, and the other used a super-linear regex on 106 distinct pages.

\myparagraph{APIs:}
We found super-linear regexes on documents associated with \SLRegexDomainCountAPI domains, spanning \SLRegexSubdomainCountAPI subdomains.

\begin{figure}[]
    \centering
    \includegraphics[width=0.95\columnwidth]{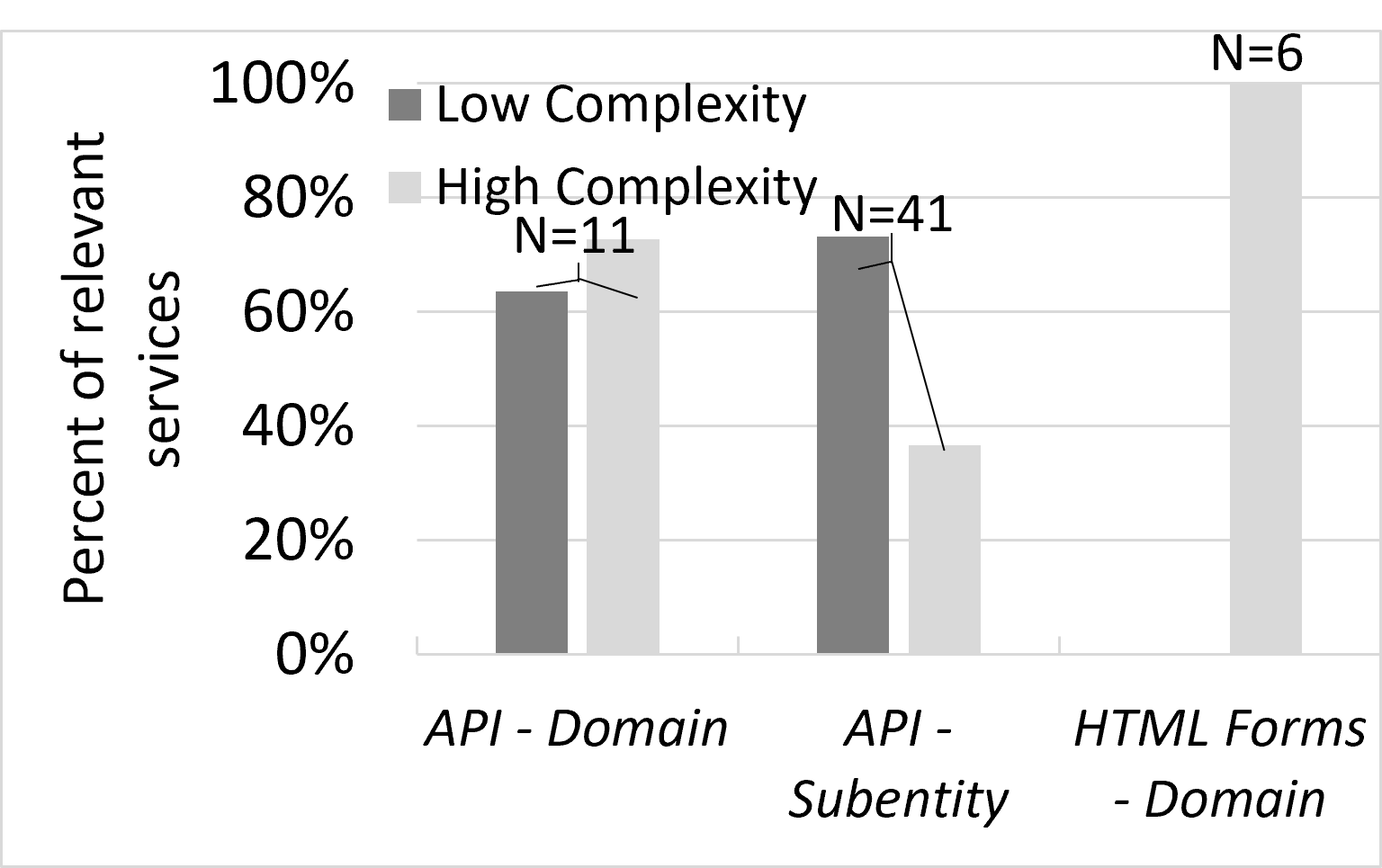}
    \caption{
      Among the web services with super-linear client-side regexes: the percent of web services with any low- and any high-complexity regexes. N refers to the total number of web services indicated by the lines pointing to the columns.
    }
    \label{fig:RegexDistribution}
\end{figure}

\myparagraph{Analysis:}
Since high-complexity regexes are more severe than low-complexity regexes, these service providers are exposed to different degrees of risk.
\cref{fig:RegexDistribution} shows the distribution of super-linear regexes by time complexity, grouped by interface type.

\begin{table*}[hbp]
\centering
\caption{
    Findings from our study of the APIs with super-linear regexes in their client-visible sanitization logic.
    Columns represent anonymized domains.
    Service provider \AWS has subdomains with varying properties. We measured response time deviations in \UnsafeDomains domains (\UnsafeSubdomains subdomains). We did not attempt to probe web services which had response time deviations indicating \REDOS caused by high time complexity regexes. 
    \emph{SL}: super-linear.
    \emph{N$\slash$A}: Domain does not have a regex of this type.
    }
\label{tab:sl-hosts-domains}
\newlength\q
\setlength\q{\dimexpr .5\textwidth -2\tabcolsep}
\noindent\begin{tabular}{cccccccc}
\toprule
\textbf{Metric}           & \textbf{\AWS} & \textbf{\Beezup} & \textbf{\Azure} & \textbf{\Tomtom} & \textbf{\Monvoyage} & \textbf{\Here} & \textbf{\ImageCharts}           \\
\midrule
Number of subdomains           & 31 & 1 & 1 & 1 & 1 & 1 & 1\\
Subdomains with SL behavior    & 10 & 1 & 1 & 0 & 1 & 1 & 1 \\
Subdomains with high-complexity SL behavior & 6 & N$\slash$A & 1 & Failed experiment & 1 & 1 & 1\\
Subdomains with low-complexity SL behavior  & 5 & 1 & Did not attempt & 0 & N$\slash$A & N$\slash$A & N$\slash$A\\
Conclusively safe subdomains                              & 13 & 0 & 0 & 0 & 0 & 0 & 0 \\
\bottomrule
\end{tabular}
\end{table*}

\subsection{RQ3: Live Unsafe Regex Engines} \label{section:Results-RQ3}
\FindingsBox{3}{
    The presence of \REDOS vulnerabilities varies widely by interface type.
    Our black-box methodology did not identify any \REDOS vulnerabilities from our analysis of HTML forms.
    From the API analysis, we identified \UnsafeDomains domains (\UnsafeSubdomains subdomains) that meet \REDOS Conditions 1--3: they apply untrusted input to a super-linear regex in an unsafe regex engine on the server side.
}

Despite our automation, the probing experiments required substantial manual intervention.
We chose to consider two kinds of equivalence classes: domains and subdomains.
Following the algorithm from~\cref{fig:oracle-flowchart}, once we reached a conclusive result in one of these equivalence classes, we did not attempt other possibilities within the class.
We did this in two distinct phases --- once at the level of domains, and once at the level of subdomains.

We probed each domain and subdomain aiming to reach a conclusive result for some super-linear regex in their interface.
We began with \SLRegexDomainCount candidate domains (\SLRegexSubdomainCount subdomains). In two domains with unsafe regexes in their HTML forms, we weren't able to obtain enough information for running probing experiments.
We identified zero \REDOS-vulnerable domains via HTML form analysis, and six \REDOS-vulnerable domains through API analysis.

\cref{tab:sl-hosts-domains} summarizes our findings for the APIs with super-linear client-side regexes.
We reached a conclusive outcome (safe or unsafe regex engine) on at least one probing experiment from \NProbedAPIDomains domains (\NProbedAPISubdomains subdomains).
The remaining five domains either did not respond to requests, or required a paid subscription.
On 5 domains (\NInconclusiveResultsAcrossAllInterfaces subdomains), at least one of our probe experiments were inconclusive.

We measured response time deviations (\ie \REDOS vulnerabilities) for at least one super-linear regex in \UnsafeDomains domains (\UnsafeSubdomains subdomains).
Notably, \SLDomainsInTopThousand of these domains are on the Tranco Top 1000 list and are major tech companies.

Following our algorithm, we concluded that there were \ConclusivelySafeSubdomains safe services by subdomain --- these subdomains have at least one super-linear client-side regex but we did not measure response time deviations on the server side.
However, 13 of these subdomains belong to domain \AWS, which also has subdomains which had measurable response time deviations.
A large company could have distinct policies at the organizational level~\cite{conway1968committees}, which may manifest by subdomain as we observed.

\myparagraph{Length-Based Mitigation:}
Davis \etal reported that input length checks are a common safety measure for low-complexity regexes~\cite{Davis2018EcosystemREDOS}.
We observed this during our experiments.
The \AWS domain had many subdomains whose only super-linear regexes were low-complexity.
Of the 13 conclusively safe subdomains of \AWS in~\cref{tab:sl-hosts-domains}, 11 used only low-complexity regexes.
The error messages from these subdomains indicated that our probe strings were too long.

\subsection{RQ4: \REDOS Mitigation} \label{section:Results-RQ4}

\FindingsBox{4}{
  The maintainers of OpenAPI middleware tools are concerned about \REDOS and interested in eliminating this possibility.
  It is unclear whether individual web service providers consider the vulnerability a threat.
}

\subsubsection{\REDOS Mitigation for OpenAPI} \label{section:RQ4-Results-OpenAPI}

All of the \REDOS vulnerabilities we identified through our black-box methodology came from APIs, not web forms.
We therefore investigated a mitigation for the OpenAPI ecosystem.
One benefit of API specifications is that client- and server-side code can be generated automatically.
In OpenAPI, two popular code generation tools are \code{swagger-codegen} and \code{openapi-generator}, in use by dozens of companies (\cref{tab:generators-validators}).
Among other features, these tools can generate server stubs, with input validation code followed by a ``fill in the blank'' for the business logic.
\emph{Their generated code  --- which includes regex checks --- can expose their dependents to \REDOS}.


\begin{table}[H]
\centering
\caption{
    Popular code generators and input validation tools for OpenAPI-based APIs.
    Data as of February 2022.
    GH Stars: the number of GitHub stars.
    }
\label{tab:generators-validators}
\newlength\qqq
\setlength\qqq{\dimexpr .5\textwidth -2\tabcolsep}
\noindent\begin{tabular}{ccc}
\toprule
\textbf{Name}           & \textbf{\# GH Stars}~\cite{Borges2018GitHubStars} & \textbf{\# Contributors}         \\
\midrule
swagger-codegen             & \SwaggerCodeGenNStars & \SwaggerCodeGenNContributors \\
openapi-generator           & \OpenAPIGeneratorNStars & \OpenAPIGeneratorNContributors \\
\bottomrule
\end{tabular}
\end{table}

These tools do not address the risk of \REDOS for their dependents.
Their documentation does not discuss how the code for regex patterns is generated.
According to our tests, these tools generate code in the target programming language and use the default (often unsafe~\cite{Davis2019LinguaFranca}) regex engine in that language.
One tool's documentation mentions the risk of \REDOS, but places the burden on the specification engineer to avoid or mitigate such regexes.
They do not allow users to tune this logic, \eg to choose a safe regex engine.

To eliminate this risk, we proposed a patch to the \code{Ajv} tool. \code{Ajv} is used by {openapi-generator} to validate requests, and has several million dependent packages according to GitHub. 

Our goal was to allow software engineers to choose the safe regex engine \code{RE2}~\cite{Cox2010RE2Implementation} instead of the built-in programming language regex engine.
\code{Ajv} is sometimes used in client-side contexts, so the engineering team prioritizes a small binary.
Adding the Node.js bindings for \code{RE2} in \code{Ajv}'s dependencies would more than double \code{Ajv}'s unpacked binary size, from 1.02MB to 2.31MB. 
Our patch therefore used the Factory pattern~\cite{gamma1995elements}, allowing users to inject another regex engine as a dependency at runtime.
The \code{Ajv} engineering team reviewed our patch and included it in release v8.8, along with documentation outlining how the patch should be used to eliminate risk of \REDOS caused by input sanitization. 
Our patch will allow the millions of \code{Ajv} dependents to eliminate this form of \REDOS in their applications.

\subsubsection{Responses to \REDOS Vulnerability Disclosures} \label{section:RQ4-Results-Disclosures}

As described in~\cref{sec:Methodology-RQ4}, we disclosed possible \REDOS vulnerabilities to live web service providers who met \REDOS conditions 1-3.

To summarize the responses:
  (1) Four service providers did not respond;
  (2) One major technology company acknowledged and repaired the disclosed vulnerability;
  (3) One major technology company initially told us that they did not perceive a vulnerability, but after several months have informed us that they have patched the unsafe regexes.
  Ultimately, both Microsoft and Amazon Web Services made changes to repair their unsafe regexes. 
Overall, the sample size is too small for comment.

\section{Discussion} \label{section:Discussion}

\myparagraph{Should web service providers prioritize \REDOS mitigations?}
Several research communities have investigated the \REDOS problem, including from 
  empirical software engineering~\cite{Davis2018EcosystemREDOS,Davis2019LinguaFranca,Davis2019RegexGeneralizability},
  systems~\cite{Cox2010RE2Implementation,Ojamaa2012NodeJSSecurity,Davis2018NodeCure}, 
  cybersecurity~\cite{Staicu2018REDOS,Davis2021SelectiveMemo},
  and
  theory~\cite{Rathnayake2014rxxr2,Weideman2016REDOSAmbiguity,Wustholz2017Rexploiter,Saarikivi2019SymbolicRegexMatcher}.
This research investment has somewhat shaky motivation:
  Crosby's proposal of \REDOS~\cite{Crosby2003REDOS},
  case studies of regex-induced service outages~\cite{StackOverflow2016REDOSPostMortem,Cloudflare2019REDOSPostMortem},
  and
  three empirical measurement studies~\cite{Wustholz2017Rexploiter,Davis2018EcosystemREDOS,Staicu2018REDOS}.
Our study provides a new perspective: large-scale black-box measurements of \REDOS risks using a weak threat model (\cref{section:Results-RQ2}, \cref{section:Results-RQ3}). Our findings establish the first systematic and empirical evaluation of \REDOS risks of web services, without an assumption of their frameworks.

Our measurements indicate that many web services are safe from \REDOS under this threat model.
In particular, for web services whose interfaces are traditional HTML forms, few sanitization regexes are revealed on the client side, and these regexes are not super-linear (\cref{section:Results-RQ1}, \cref{section:Results-RQ2}).
In contrast, web services that publish APIs face more risk of \REDOS.
In publishing API specifications, web services are choosing to reveal more about their server-side sanitization logic.
However, the cause is unclear.
We conjecture two explanations for further examination.
First, the choice may be deliberate.
Software engineers may be providing a fuller definition of their input validation constructs in their API specifications, so that code generation tools can be used to automatically handle input validation. 
Alternatively, it may be accidental.
Software engineers may be using tools which generate API specifications from code (similar to model extraction), rather than writing specifications first and then generating code from them.
This process may be inadvertently exposing internally-used regexes, which could explain the greater regex variety and greater incidence of \REDOS among API regexes.

\myparagraph{A visibility-security tradeoff:}
Our measurements indicate a tradeoff between visibility and security (\cref{section:Results-RQ3}).
Web service providers who promote usability by specifying the nature of valid input may expose themselves to \REDOS.
Although this class of attacks could be mitigated by hiding the sanitization rules, software engineers should not seek security through obscurity~\cite{SecurityThroughObscurity}.
Indeed, describing the characteristics of valid input is necessary to enable communication.
Software engineers should not need to choose between visibility and security.

Rather than obscuring input formats, the software engineering community would benefit from principled solutions to \REDOS.
Davis \etal described several sound solutions, and they concluded that making regex engines safe seemed like the most natural mitigation~\cite{Davis2021SelectiveMemo}.
Further research into adopting safe regex engines~\cite{Cox2010RE2Implementation,Wang2019IntelHyperscan} or retrofitting existing unsafe engines~\cite{Davis2021SelectiveMemo} will improve the safety of software engineering practice.
Middleware for input sanitization can use a level of indirection to select a safe regex engine.
In our mitigation study (\cref{sec:Methodology-RQ4}), we found that middleware providers were happy to accept such a change.
However, they were concerned with introducing an external dependency on a safe regex engine, indicating that improving the safety of programming language regex engines should be an area of focus for a long-term solution.


Using API specifications has many advantages from a software engineering standpoint.
Specific to \REDOS, the standardization they provide allows engineers to use specification-compatible middleware tools.
Hence, any security patches to these tools can protect many web services at once.
We took advantage of this centralization to provide ReDoS mitigations to the numerous dependents of such middleware tools.
The same property that lends itself to easy exploitation also lends itself to a centralized repair.

\myparagraph{Mismatch between OpenAPI SDL and needs in practice:}
Ideally, an interface specification should describe everything necessary for successful communication.
However, during our measurements we observed that most OpenAPI documents were underspecified.
Our requests, which passed the specified validation constraints according to the Prism tool, were still invalid according to the server.

We identified several causes of these underspecifications.
One of these causes is well known --- the OpenAPI syntax cannot indicate dependencies between requests.
Although RESTler~\cite{Atlidakis2019RESTler} attempts to infer these dependencies using heuristics, it cannot handle all cases.
We encountered several more causes during our experiments.
Some API documents indicate the full set of possible payload variables, but actually accept subsets of those variables.
Some parameters are interdependent/coupled, \eg including one optional parameter requires including others~\cite{openapiDependencyIssue}.
Some parameters of ``\code{string}'' type are actually type-aliased within the web service, \eg strings that represent a comma-separated numeric sequence.
These various missing semantics are of practical utility, and the maintainers of API specification languages should consider supporting them.

\section{Related Work} \label{section:RelatedWork}

Our work descends from two lines of research:
  web service vulnerability scanning, and regular expression engineering.

We employed a black-box probing methodology to scan web services for a specific class of security vulnerability.
Other researchers have employed grey-box and white-box methodologies for this vulnerability~\cite{Wustholz2017Rexploiter,Staicu2018REDOS}.
Researchers and commercial tools offer black-box and grey-box scanning for diverse vulnerabilities~\cite{Bau2010BlackBoxTesting,Doupe2012EnemyOfTheState,Dogan2014},
  including
  algorithmic complexity vulnerabilities~\cite{Li2021Markdown,Petsios2017SlowFuzz,Cha2020GraphQL},
  service crashes~\cite{Atlidakis2019RESTler,Godefroid2020RESTlerDataFuzzing,Atlidakis2020RESTlerSecurityProperties},
  cross-site scripting (XSS)~\cite{Bates2010XSSRegexFilter,Tripp2013Learning,Duchene2014XSS},
  and SQL injection~\cite{Martin2008GeneratingXSSAndSQL}.
Notably, some researchers have pursued the opposite of our Consistent Sanitization Assumption to identify cases where backend sanitization appears problematically inconsistent~\cite{Offutt2004BypassTesting,Li2010Perturbation}.

While our study focuses on regex cybersecurity, researchers have considered other aspects of the regex engineering lifecycle.
Michael \etal reported that many software engineers find regex engineering difficult~\cite{Michael2019RegexesAreHard}.
To assist the engineering community in this domain,
  researchers have recently
    described regex engineering practices related to composition~\cite{Bai2019RegexComposition}, comprehension~\cite{Chapman2017RegexComprehension}, and testing~\cite{Wang2018RegexTestCoverage};
    identified common regex bug patterns and taxonomies~\cite{Larson2016EvilTestStrings,Wang2020RegexBugs,Eghbali2020StringBugs};
    and
    proposed tools to support regex comprehension~\cite{Beck2014RegViz}, testing~\cite{Spishak2012TypedRegexes,Larson2018AutomaticCheckingOfRegexes}, and repair~\cite{Pan2019AutomaticRegexRepair,Li2020REDOSRepair}.
There has also been a longstanding effort to automatically compose regexes, with diverse approaches including 
  formal methods~\cite{Angluin1978RegexGenerationNPHard,Gold1978RegexGenerationNPHard,Denis2001LearningRegexes,Alquezar1999LearningRegexes,Chida2020REDOSRepair,Li2020RegexSynthesis},
  evolutionary algorithms~\cite{Bartoli2012RegexSynthesis,Bartoli2016EvolvingRegexes,CodyKenny2017RegexGeneticPerf},
  optimization~\cite{Li2008LearningRegexes,Rebele2018GeneralizingRegexes},
  crowd-sourcing~\cite{Cochran2015CrowdsourcedRegexSynthesis},
  natural-language translation~\cite{Chen2020MultiModalRegexSynthesis},
  and human-in-the-loop interactive development~\cite{Drosos2020Wrex,Zhang2020InteractiveSynthesis}.



\section{Threats to Validity} \label{section:Threats}

This paper describes a substantial web measurement study covering two distinct interface types.
We acknowledge a variety of threats to the validity of our findings, and note mitigating factors.

\myparagraph{Construct validity:}
The primary threat here is in our definition of a browser-based web interface: via HTML forms.
While HTML forms appeared in \WebFormRegexDomainPercentHasForm of the web services whose HTML form interfaces we crawled, trends such as Single-Page Applications~\cite{MDNDocsSinglePageApplication} process user input without form submissions.
This is also a threat to external validity, as we cannot comment on the risk of \REDOS for such web services.
Beyond this construct, we relied on definitions of super-linear regexes and regex-based denial of service.
These concepts are well established in the research literature, and we measured them with state-of-the-art tools and probing methodologies.

\myparagraph{Internal validity:}
Our methodology does not let us measure the degree of a \REDOS vulnerability.
We identified super-linear regexes that are applied to user input in an unsafe regex engine on the server side.
However, we cannot assess \REDOS Condition 4 (\REDOS mitigations~\cref{section:Background}), without either having server-side knowledge or launching a full-scale denial of service attack.
To shed light on this threat, in~\cref{section:Results-RQ4} we discussed perspectives from the web service engineering community.

There are three potential sources of under-reporting in our probing methodology.
First, as noted in~\cref{sec:Methodology-RQ3}, some web services cache end-to-end results, and this caching will mask worst-case behavior when we use identical worst-case probe strings.
Second, we conservatively chose input lengths for our probes based on the slowdowns observed on a workstation-grade machine.
If a web service provider processes input on a server-class machine, the response time deviation induced by our probes may not be observable.
Finally, our decision tree yielded inconclusive results in \NInconclusiveResultsAcrossAllInterfaces cases (\cref{fig:oracle-flowchart}).

\myparagraph{External validity:}
Our goal was to measure the extent of \REDOS vulnerabilities in live web services.
The populations we used may have biased our results.
We probed web services that were listed in directories of live web services --- from the Tranco Top 1M directory~\cite{pochattranco2019} for HTML form interfaces, and from \code{apis.guru} for OpenAPI interfaces.
For \emph{HTML forms}, we randomly sampled from the top \WebFormsTopXPopulationSize web domains for HTML forms. 
We expect these results to generalize to other popular websites; popularity may be correlated with a certain caliber of engineering (in order to service the client load) and so our results may not generalize to less popular websites.
A related bias is that \PercentServicesRejectingGCPConnections of HTML form services rejected our connections outright because we were using VMs from Google Cloud Platform for our experiments.
For \emph{APIs}, we considered all API specifications from \code{apis.guru}.
This directory only contains OpenAPI specifications from \ApisGuruDomainCount distinct domains (\ApisGuruSubdomainCount distinct subdomains).
Different results may emerge from studying other API specification directories, \eg SwaggerHub or by mining GitHub.
We performed a preliminary analysis on SwaggerHub specifications, and found that often they are simpler and are not associated with a live web service URL.
Echoing Wittern \etal~\cite{Wittern2019EmpiricalGraphQL}, we suggest that the results we obtained from \code{apis.guru} may be more representative of \emph{engineered} OpenAPI specifications.

Beyond limitations in our sampling, generalizability is threatened by our black-box methodology.
Our approach depends on the \emph{Consistent Sanitization Assumption}: that web services are consistent in their input sanitization, \ie that client-visible input sanitization is also applied on the back-end.
This assumption permits a scalable black-box approach.
However, without full knowledge of server-side logic, we may omit server-side regex evaluations that are not exposed to clients.
Web services may apply additional or alternative sanitization on the back-end.
For example, the \REDOS vulnerabilities identified by Staicu \& Pradel~\cite{Staicu2018REDOS} would likely not have been discovered using our methodology, since they targeted regexes that would only be used server-side in HTTP header processing.
Conversely, client-side sanitization gives us insight into ``business logic'' regexes that might not be visible through examination only of the open-source software supply chain, as Davis \etal~\cite{Davis2018EcosystemREDOS} and Staicu~\&~ Pradel~\cite{Staicu2018REDOS} did.
Our findings thus complement the prior empirical studies of the risks of \REDOS in practice.



\section{Future Work} \label{section:Future Work}

\myparagraph{Transferring \REDOS vulnerabilities:}
Software engineers solve similar problems in similar ways~\cite{knightExperimentalEvaluationAssumption1986}.
A general version of the Consistent Sanitization Assumption is possible: that \emph{web services validate similar content in similar ways}, so sanitization logic revealed by one service may be transferred to another.
For example, suppose two web services use an accessibility feature like ARIA labels~\cite{W3Aria} to label a form field as an email.
If one service provides client-side sanitization logic, similar logic might be in use by the other.

\myparagraph{Why are API practices more dangerous?}
In answering each research question, there were marked differences between \REDOS risks in traditional HTML forms as compared to the emerging approach of API specification.
We conjectured two causes (\cref{section:Discussion}):
   providing detailed API specifications to ease the development of input validation logic,
   and
   inadvertent exposure resulting from API extraction from server-side code.
We believe this finding bears further investigation.

\myparagraph{Improved tooling:}
Although we chose RESTler to help us reach endpoints in complex APIs, we eventually performed manual intervention for most of the \NProbedAPISubdomains APIs we probed.
 In practice, API specifications underspecify valid interactions.
When we intervened, we consulted API documentation as well as the service error messages.
Incorporating NLP techniques into automated API interactions is a natural direction for improved black-box web service testing~\cite{Godefroid2020RESTlerDataFuzzing}.

\myparagraph{Regex dataset:}
Previous researchers have collected regex datasets from open-source software repositories~\cite{Chapman2016RegexUsageInPythonApps,Davis2019RegexGeneralizability}, with applications including improved regex usability tools and safer regex engines~\cite{Turonova2020CountingSetAutomata,Davis2021SelectiveMemo}.
To complement this effort, we contribute a dataset of \emph{web input sanitization} regexes.
This dataset contains the $\sim1850$ unique regexes identified during our experiments.


\section{Conclusions} \label{section:Conclusions}

Regex-based denial of service (\REDOS) has received much recent attention.
Web service providers are curious about the degree to which \REDOS threatens them, and regex engine maintainers wonder whether they should prioritize optimizations to ameliorate \REDOS.
In light of this interest, we report the results of the first black-box measurement study of \REDOS vulnerabilities on live web services.
Our method is based in the observation that server-side input sanitization may be mirrored on the client-side as part of usability engineering.
We therefore examined the extent to which super-linear regexes on the client side can be exploited as \REDOS vulnerabilities on the server side.
We compared two common interface types: HTML forms ($N=\WebFormsSampleSize$ domains) and APIs ($N=\ApisGuruDomainCount$ domains). 
We report that although client-visible regexes are common in both types of interfaces, super-linear regexes are only common in APIs.
We identified \REDOS vulnerabilities in the APIs of \UnsafeDomains domains (\UnsafeSubdomains subdomains) including in services operated by major technology companies.
Our findings add weight to the concerns of researchers about the risks of \REDOS in practice.
Specifically, we show that the movement toward API specification development provides leverage for ReDoS attacks.


\section*{Acknowledgments} \label{section:Acknowledgments}

We thank A. Kazerouni and the anonymous referees for their criticisms.
Barlas and Du acknowledge support from Purdue University's Summer Undergraduate Research Fellowship program (SURF) and the Purdue University Center for Programming Principles and Software Systems (PurPL).

\section*{Research ethics} \label{section:Ethics}

Our methodology could be construed as conducting denial of service attacks against live web services.
Attacking web services is unethical.
We did not do so!
As discussed in~\cref{section:Methodology}, we imitated Staicu \& Pradel~\cite{Staicu2018REDOS} by using \emph{\REDOS probes} instead of attacks.
Prior researchers have contributed theoretical and empirical understanding of worst-case regex performance, enabling us to accurately predict the worst-case performance of a problematic regex.
This prediction allows us to size the probes to minimize the risk to the service provider.
Our method could introduce a user-perceivable slowdown comparable to a network hiccup, yet still demonstrates the possibility of a \REDOS attack by a malicious actor.
We believe this cost is an acceptable price for the data we have collected.
However, as a consequence of our ethical probing methodology, we are limited in what we can claim about the vulnerability of web services (\cref{section:Discussion}).

\section*{Data Availability} \label{section:Reproducibility}
An artifact is available at 
\url{https://doi.org/10.5281/zenodo.5916441}.
The artifact has a dataset and our vulnerability identification tools.
\emph{Dataset}: We provide a list of regexes found in web forms and API specifications and their analysis reports per \code{vuln-regex-detector}.
We also share the list of web services with regexes in their API specifications, and the list of all web services with API specifications.

\emph{Vulnerability identification tools}: We share one tool for browser-based interfaces, and one tool for APIs (OpenAPI).
The browser-focused tool crawls web services for regex use in web forms and JavaScript files.
It produces OpenAPI specifications for endpoints which are using vulnerable regexes in the front-end.
The API-focused tool parses OpenAPI specifications for vulnerable regexes and endpoints.
It also probes the web services to assess whether each vulnerable regex is exploitable for ReDoS.
\appendix

\raggedbottom
\pagebreak


\bibliographystyle{ACM-Reference-Format}
\bibliography{./bibliography/ALLREFS}
\end{document}